\documentclass[conference]{IEEEtran}

\usepackage{epsfig}
\usepackage{graphicx}
\usepackage{amssymb}
\usepackage[noend]{algorithmic}
\usepackage{algorithm}
\usepackage{algorithmext}
\usepackage{multirow}
\usepackage{subfigure}
\usepackage[cmex10]{amsmath}
\usepackage{balance}
\usepackage{sidecap}

\begin{document}
%

\title{High-throughput Execution of Hierarchical Analysis Pipelines on Hybrid Cluster Platforms}

\author{
George Teodoro, Tony Pan, Tahsin M. Kurc, Jun Kong, Lee A. D. Cooper, and Joel H. Saltz\\
\IEEEauthorblockA{Center for Comprehensive Informatics, Emory University, Atlanta, GA 30322\\
Email:\{george.teodoro,tony.pan,tkurc,jun.kong,lee.cooper,jhsaltz\}@emory.edu }

}
\maketitle
\begin{abstract}
We propose, implement, and experimentally evaluate a runtime middleware to support 
high-throughput execution on {\em hybrid} cluster machines of large-scale 
analysis applications. A hybrid cluster machine consists of computation nodes which 
have multiple CPUs and general purpose graphics processing units (GPUs). Our work 
targets scientific analysis applications in which datasets are processed in 
application-specific data chunks, and the processing of a data chunk is expressed 
as a hierarchical pipeline of operations. The proposed middleware system combines 
a bag-of-tasks style execution with 
coarse-grain dataflow execution. Data chunks and associated data processing pipelines 
are scheduled across cluster nodes using a demand driven approach, while within a node 
operations in a given pipeline instance are scheduled across CPUs and GPUs. 
The runtime system implements several optimizations, including performance aware 
task scheduling, architecture aware process placement, data locality conscious task assignment, 
and data prefetching and asynchronous data copy, to maximize utilization of the aggregate 
computing power of CPUs and GPUs and minimize data copy overheads. The application 
and performance benefits of the runtime middleware are demonstrated using an image 
analysis application, which is employed in a brain cancer study, on a state-of-the-art 
hybrid cluster in which each node has two 6-core CPUs and three GPUs.
Our results show that implementing and scheduling 
application data processing as a set of fine-grain operations provide more 
opportunities for runtime optimizations and 
attain better performance than a coarser-grain, monolithic implementation. 
The proposed runtime system can achieve high-throughput processing of large 
datasets -- we were able to process an image dataset consisting of 36,848 
4Kx4K-pixel image tiles at about 150 tiles/second rate on 100 nodes. 
\end{abstract}

\section{Introduction} \label{sec:intro}
The processing power and memory capacity of graphics processing units (GPUs)
have rapidly and significantly improved in recent years. Contemporary GPUs
provide extremely fast memories and massive multi-processing capabilities,
exceeding those of multi-core CPUs. The application and performance benefits 
of GPUs for general purpose processing have been demonstrated for a wide range 
of applications~\cite{cudaapps}. As a result, 
CPU-GPU equipped machines are emerging as viable high performance 
computing platforms for scientific computation~\cite{10.1109/MCSE.2011.83}. 
More and more supercomputing 
systems are being built with {\em hybrid} computing nodes that have multi-core 
CPUs and multiple GPUs. This trend is also fueled by the availability of 
programming abstractions and frameworks, such as
CUDA~\cite{cuda} and OpenCL~\cite{opencl08}, that have reduced the complexity 
of porting computational kernels to GPUs. Nevertheless, taking advantage of hybrid 
platforms for scientific computing still remains a challenging problem. An application 
developer 
needs to deal with the efficient distribution of computational workload not only across 
cluster nodes but also among multiple CPU cores and GPUs, which have different performance 
characteristics and memory capacities, on a hybrid node. The developer 
also has to take into account potential performance variability across application 
operations. Operations ported to the GPU will not all have the same amount of 
performance gains. Some operations are more suitable for massive parallelism and 
generally achieve higher GPU-vs-CPU speedups than other operations. In addition, the 
application developer has to minimize data copy overheads when data have to be 
exchanged between application operations. These challenges often lead to 
underutilization of the power of hybrid platforms.   

In this work we investigate the design and implementation of runtime middleware
support to address these issues in the context of large-scale scientific data 
analysis applications. 

Analysis of large datasets is a critical, yet challenging component of
scientific studies, because of dataset sizes and the computational requirements
of analysis applications. Sophisticated sensors enable 
scientists in biomedicine and earth systems sciences to perform high resolution 
measurements of objects under study rapidly. Similarly, leadership scale machines 
at national laboratories and supercomputer centers have made it possible for  
researchers to carry out large scale simulations of
complex physical phenomena and generate terabytes of
data per simulation run. Processing a large dataset can take very long time on 
even high end workstations. Moreover, a dataset may be analyzed multiple times with 
different analysis parameters and algorithms to explore different
scientific questions, to carry out sensitivity studies, and to quantify
uncertainty and errors in analysis results.

While the types of operations and algorithms employed by a scientific project
for data analysis will be specific to the objectives of the project, scientific
data analysis applications exhibit common data access and processing patterns.
Large datasets can often be processed in a set of {\em data chunks}, where each
chunk represents an application-specific (or user-defined) portion of the
dataset; in a large image dataset, for instance, each image or an image tile
can be a data chunk. Common processing patterns include bag-of-tasks
execution~\cite{Carriero:1990:WPP:98120}, generalized reduction and MapReduce
patterns~\cite{dean04mapreduce}, and coarse-grain dataflow
patterns~\cite{beynon01datacutter,Eisenhauer00eventservices,Abbasi:2009:DSD:1551609.1551618,DBLP:conf/ppopp/SubhlokSOG93,anthillII,Teodoro:2008:RSE:1993317.1993323,10.1109/CCGRID.2007.20}.
In a bag-of-task style execution, application-specific (or user-defined) chunks
of a dataset are processed concurrently. MapReduce has gained popularity in
recent years as a framework to support large scale data processing that can be
expressed as Map and Reduce operations. Some analysis 
methods, on the other hand, are more appropriately expressed and executed as 
a pipeline of operations. For instance, segmentation of a nucleus consists of 
several steps including creation of image masks, shape creation and morphing, and
determination of boundaries. This type of processing can be described more
naturally using a coarse-grain dataflow (or filter-stream) pattern~\cite{beynon01datacutter}, 
in which application processing is carried out as a network of components connected
through logical pipes. Each component performs a portion of the
application-specific processing, and interactions between the components are
realized by flow of data and control information. 

In prior work, Mars~\cite{mars} and Merge~\cite{merge} evaluated the
cooperative use of CPUs and GPUs to speedup MapReduce computations. Mars
performed an initial evaluation on the benefits of partitioning Map and Reduce
tasks between CPU and GPU statically.  Merge extended that approach with
dynamic distribution of work at runtime.  The Qilin~\cite{qilin09luk} system
further proposed an automated methodology to map computation tasks to CPUs and
GPUs. Unfortunately, neither of these solutions (Mars, Merge, and Qilin) are
able to take advantage of distributed systems. Some projects more recently
focused on execution in distributed CPU-GPU equipped
platforms~\cite{6061070,ravi2010compiler,HartleySC10,6152715,hpdc10george,Phillips:2008:AMP:1413370.1413379,Jetley:2010:SHN:1884643.1884689,cluster09george,springerlink:10.1007/s10586-010-0151-6}.
Ravi et al.~\cite{ravi2010compiler,6152715} proposed techniques for automatic
translation of generalized reductions to CPU-GPU environments via compiling
techniques, which are coupled with runtime support to coordinate execution. The
runtime system techniques introduced a number of auto-tuning approaches to
partition tasks among CPUs and GPUs for generalized reduction operations. The
work developed by Hartley et al.~\cite{HartleySC10} is contemporary to that of
Ravi and proposed similar runtime approaches to divisible workloads. The work 
by Bosilca et al.~\cite{6061070} presented DAGuE, a framework to enable the use of
heterogeneous accelerated machines to executed dense linear 
algebra operations. 

Our 
solution combines the coarse-grain dataflow pattern with the 
bag-of-tasks pattern in order to facilitate 
the implementation of an analysis application from a set of processing 
components. It supports hierarchical pipelines, in which a processing component can itself
be a pipeline of operations, and implements optimizations 
for efficient use of CPUs and GPUs in coordination on a computing node. 
The runtime optimizations include 
data locality conscious and performance variation aware task assignment, 
data prefetching, asynchronous data copy, and architecture aware placement of 
control processes in a computation node. 
Fine-grain 
operations that constitute an analysis pipeline typically involve different 
data access and processing patterns. Consequently,
variability in the amount of GPU acceleration of operations is
likely to exist. This requires the use of performance aware scheduling
techniques in order to optimize the use of 
CPUs and GPUs based on speedups attained by each operation. In addition, 
our middleware automatically applies  data locality conscious task assignment 
to ensure good performance even in the
case where performance variability is not present or speedup 
estimates are not available. Data prefetching and asynchronous data transfers 
between CPUs and GPUs 
are also employed in order to maximizes the GPU utilization and reduce data 
copy overheads by enabling data flow between devices in parallel to ongoing 
computation. We demonstrate the application of the 
runtime system via the
implementation of a biomedical image analysis application, used in
study of brain tumors. We carry out a performance 
evaluation of the runtime system  on a state-of-the-art distributed 
memory cluster machine where each node has two 6-core CPUs 
and 3 high-end GPUs.

\section{Application Scenario} \label{sec:app}
\begin{figure*}
\begin{center}
\includegraphics[width=0.9\textwidth]{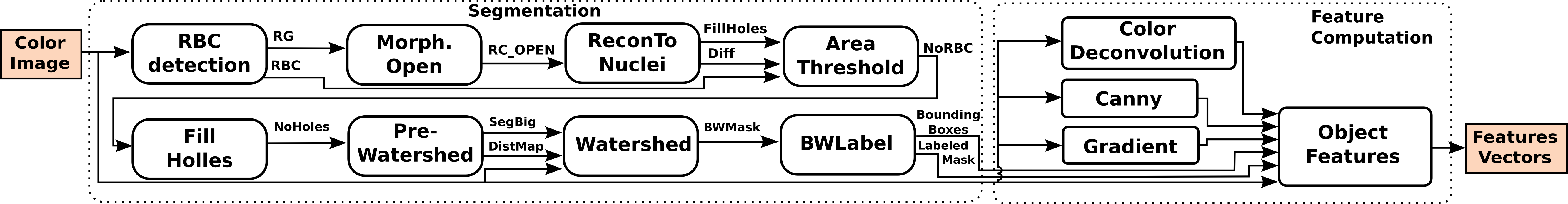}
\vspace*{-2ex}
\caption{Pipeline for segmenting nuclei in a whole slide tissue image and computing their 
features. The input to the pipeline is an image or image tile. The output is a set of 
features for each segmented nucleus.}
\label{fig:pipeline}
\end{center}
\vspace*{-6ex}
\end{figure*}

Biomedical research studies that make use of large datasets of digital microscopy 
images are a good example of scientific applications targeted in our work. An example 
of such studies is the work done at the In Silico Brain Tumor Research Center 
(ISBTRC)~\cite{5518399,insilico}. ISBTRC conducts research on brain tumors, to find better tumor 
classification strategies and to understand the biology of brain tumors, using 
complementary datasets of high-resolution whole tissue slide images (WSIs), gene 
expression data, clinical data, and radiology images. As part of this
research effort, our group has developed image analysis applications 
to extract and classify morphology and texture information from high resolution WSIs, 
with the objective of exploring correlations between tissue morphology
features, genomic signatures, and clinical data~\cite{jamia-insilico,ieee-insilico}. 
The WSI analysis applications share 
a common set of cascaded stages, including: 1) image preprocessing tasks such as color
normalization, 2) segmentation of micro-anatomic objects such as cells and nuclei, 
3) characterization of the shape and texture
features of the segmented objects, and 4) machine-learning methods that integrate
information from features to classify the objects.  
In terms of computation cost, the preprocessing and classification stages 
(stages 1 and 4) are
inexpensive relative to the segmentation and feature computation stages 
(stages~2~and~3). The classification stage includes
significant data reduction prior to the actual classification operation which 
reduces computational requirements. The segmentation and feature computation 
stages, on the other hand, may operate on hundreds of images with 
resolutions ranging from $50K$x$50K$ to $100K$x$100K$ pixels and $10^{5}$ to 
$10^{7}$ micro-anatomic objects (e.g., cells and nuclei) per image. 
Our optimization efforts to date, therefore, have been focused 
on these two stages.

The segmentation stage detects cells and nuclei and delineates their boundaries. 
It consists of several component operations, forming a coarse-grain dataflow graph 
(see Figure~\ref{fig:pipeline}). 
The operations include morphological reconstruction to identify candidate objects, 
watershed segmentation to separate 
overlapping objects, and filtering to eliminate candidates that are unlikely to be 
nuclei based on object characteristics. The feature computation stage derives quantitative 
attributes in the form of a feature vector for the entire image or for individual segmented 
objects. The feature types include pixel statistics, gradient statistics, Haralick features~\cite{1455597}, edge, and morphometry. Most of the features 
can be computed concurrently in a multi-threaded or parallel environment.

This application scenario encapsulates several processing patterns. First, each
image can be partitioned into rectangular tiles, and the pre-processing,
segmentation, and feature computation stages can be executed on each tile
independently. This leads to a bag-of-tasks style processing pattern.
Similarly, feature computations for individual objects can also be executed in
this pattern. Second, the processing of a single tile is expressed as a
hierarchical coarse-grain dataflow pattern. The pre-processing, segmentation,
and feature computation stages are the first level of the dataflow structure.
The segmentation stage itself consists of a pipeline of operations. Third, the
classification stage can be expressed as a MapReduce computation, in which
feature vectors for individual objects are aggregated to form average feature
vectors per image and per patient. These average feature vectors are then used
in machine-learning algorithms, such as k-means~\cite{kmeans}, to classify
patients and images into groups. 

In this paper we target the segmentation and feature computation stages.  As
part of our overall effort to scale the application, we also have been
developing CPU and GPU versions of the components in the segmentation stage as
well as those in the feature computation stage. We have used CPU and GPU
implementations from the OpenCV library~\cite{opencv_library} or from other
research groups such as in the case of the watershed
segmentation~\cite{Korbes:2011:AWP:2023043.2023072}.  For those operations for
which we could not find existing implementations, we have developed our own
implementations. Table~\ref{tab:pipeline-ops} in Section~\ref{sec:results}
summarizes the source of the CPU/GPU implementations. Having CPU and GPU
versions of the data processing components allows the runtime system to utilize
CPU cores and GPUs on a computation node concurrently in a coordinated way, as
shall be described in the next sections.

\section{Middleware Runtime Framework} 
\subsection{Application Representation}
The application representation model draws from DataCutter~\cite{beynon01datacutter}, 
which is a filter-stream middleware framework. An analysis application is represented as 
a pipeline of operations. The application operations are connected through logical 
streams; an operation reads data from one or more streams, processes the data, and 
writes the results to one or more streams. We have adapted the DataCutter application 
model in our framework in the following ways. 

\begin{figure}[ht]
\begin{center}
\includegraphics[width=0.46\textwidth]{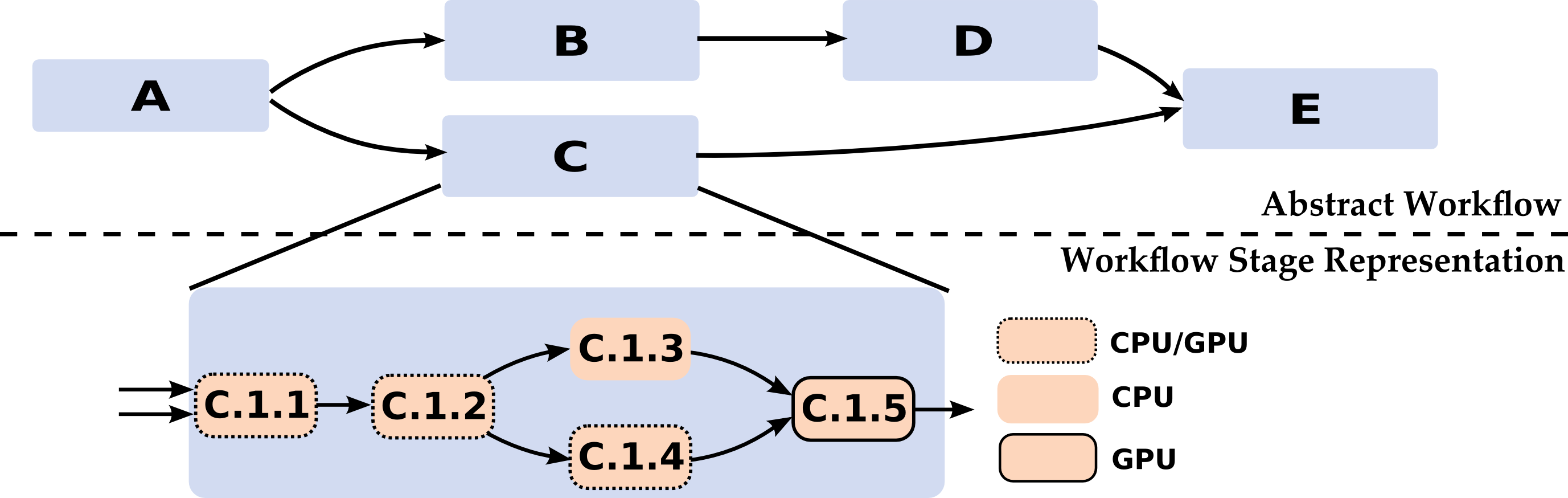}
\vspace*{-2ex}
\caption{Hierarchical pipeline model.}
\label{fig:sampleDataflow}
\end{center}
\vspace*{-3ex}
\end{figure}

The new framework supports hierarchical pipelines in that an operation
can itself be made up of a pipeline of lower-level operations. We will describe 
the framework in the context of two pipeline levels for the sake of presentation, 
although the framework allows for multiple levels of hierarchies. 
The first level is the {\em coarse-grain operations}
level, which represents the main stages of an analysis application. The {\em
fine-grain operations} level is the second level and represents lower-level
operations, from which a main stage is created. 
Figure~\ref{fig:sampleDataflow} illustrates the hierarchical pipeline
representation of an analysis application. The top of the figure shows the
coarse-grain operations, while the fine-grain level is displayed in the bottom
portion of the figure. In the
example image analysis application, for instance, the segmentation and feature 
computation stages
constitute the first level, whereas individual operations in those stages are 
represented in the second level.

\begin{figure}[ht]
\begin{center}
\includegraphics[width=0.46\textwidth]{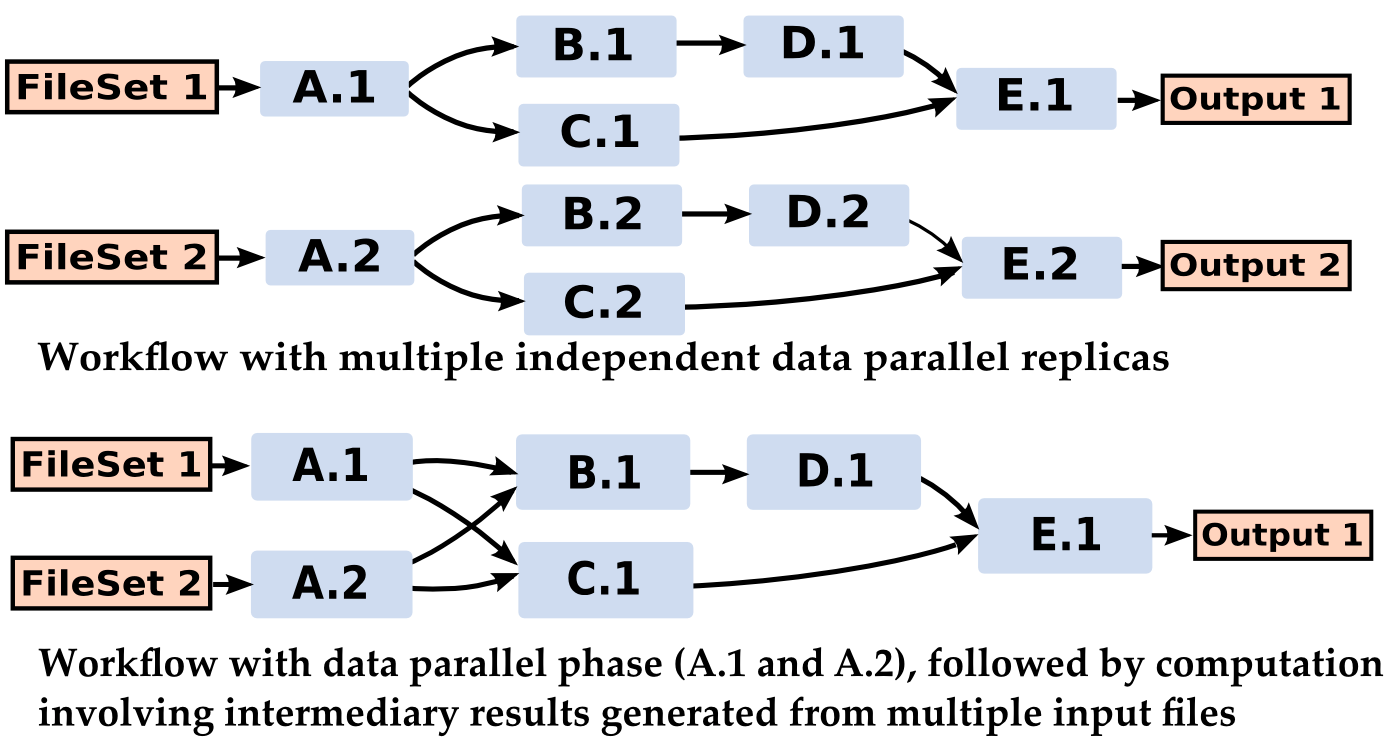}
\vspace*{-2ex}
\caption{Examples of {\em Concrete Workflow} instantiations.}
\label{fig:concreteWorkflow}
\end{center}
\vspace*{-3ex}
\end{figure}

The framework distinguishes between two representations of a pipeline. The
{\em Abstract Workflow} representation describes the logical stages of the
analysis application and the connections, or dependencies, between the stages.
The {\em Concrete Workflow} representation, on the other hand, is a binding of
the logical stages and operations to actual operations and input data. The
concrete workflow is in essence an instantiation of the abstract workflow
representation. A \emph{stage instance} is represented by a tuple, 
\emph{(input data chunk, processing stage)}; similarly, an \emph{operation instance} 
is represented by a tuple, \emph{(input data chunk, operation)}. When a stage or
operation is instantiated, the dependencies that are expressed in the abstract
workflow are exported to the runtime environment for correct execution. Two
examples of {\em Concrete Workflow} instantiations of the {\em Abstract Workflow} in
Figure~\ref{fig:sampleDataflow} are presented in
Figure~\ref{fig:concreteWorkflow}.

The framework makes use of the concept of function variants to leverage CPUs
and GPUs in the computing system. A function variant is a group of functions
with same name, arguments, and result
types~\cite{Millstein:2004:PPD:1035292.1029006,merge}. When a logical stage or
operation is bound to a concrete operation, the concrete operation can be a
single function or a function variant. In our implementation a function variant
for a data processing operation is the CPU and GPU implementations of the
operation. Binding to a function variant enables the runtime system to choose
the appropriate function or functions during execution, allowing multiple
computing devices to be used concurrently and in a coordinated manner. 
Note that the concept of function variants is more
comprehensive, and several variants of an operation for the same computing device
can coexist.
\subsection{Runtime System Implementation} \label{sec:sysimpl}
The hierarchical representation lends itself to a separation of concerns and
enables the use of different scheduling approaches at each level. For instance,
it allows for the possibility of exporting second level operations ({\em
fine-grain operations}) to a local scheduler on a hybrid node, as opposed to
describing each pipeline stage as a single monolithic task, while employing a
different mapping and scheduling approach for the first level. In this way, the
scheduler can control tasks in a smaller granularity and can account for
performance variations across the finer grain tasks within a node, while
reducing the scheduling overhead for utilization of computation nodes across
the machine. 

\begin{figure}[ht]
\begin{center}
\includegraphics[width=0.49\textwidth]{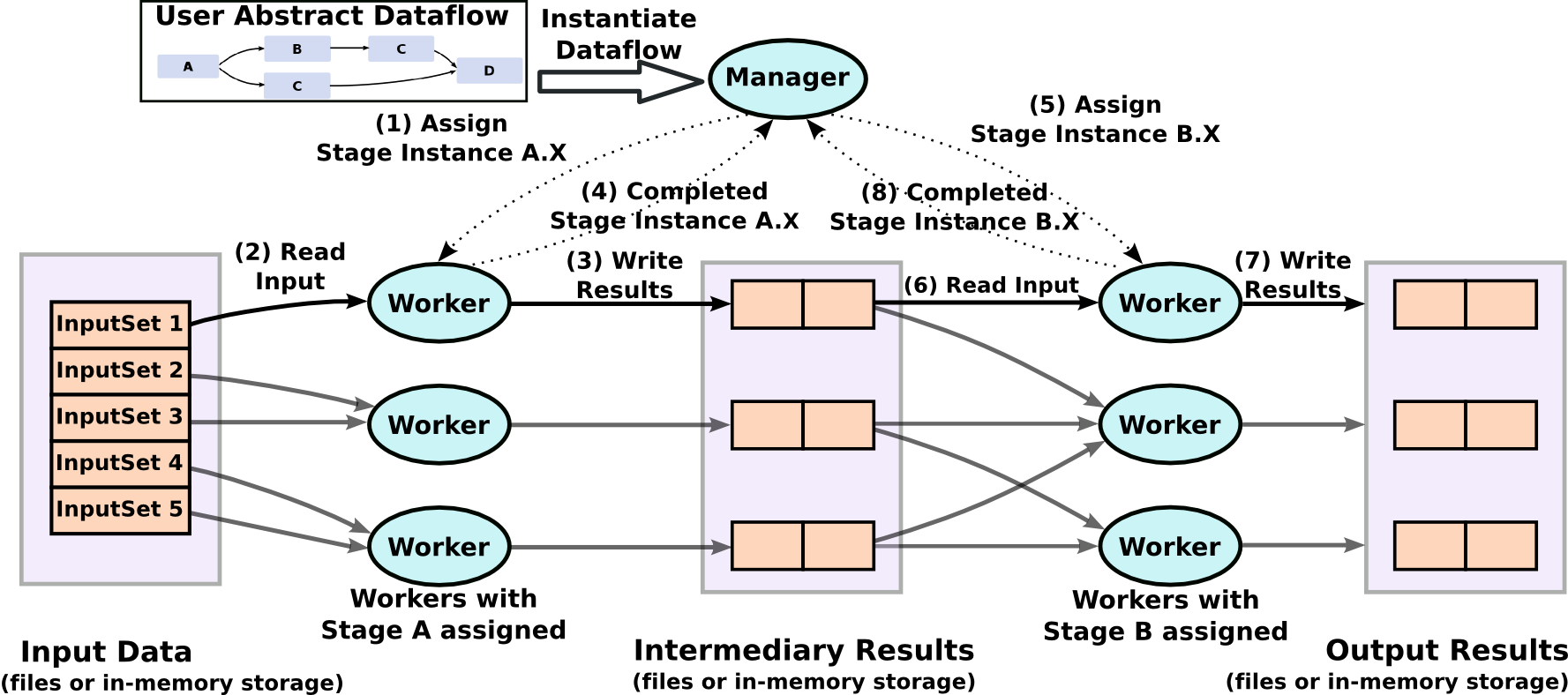}
\vspace*{-5ex}
\caption{Overview of the middleware system architecture and task scheduling.}
\label{fig:execModel}
\end{center}
\vspace*{-2ex}
\end{figure}

In our current implementation, the runtime system uses a Manager-Worker model,
as shown in Figure~\ref{fig:execModel}, in order to combine the bag-of-tasks
style execution with the coarse-grain dataflow execution pattern. 

The Manager has an overall view of the runtime environment and is responsible
for instantiating an abstract workflow and tracking dependencies between the
workflow stages to ensure correct execution. An abstract workflow can be
instantiated in several ways to take advantage of distributed computing power
on a cluster system. If a dataset is divided into data chunks (e.g., image
tiles in an image dataset) and each tile can be processed independently, the
entire pipeline can be replicated across the system. Each replica is assigned
to one or a group of nodes and processes a subset of data chunks. This type of
instantiation is shown at the top of Figure~\ref{fig:concreteWorkflow}.  An
alternative approach is to instantiate and execute different numbers of copies
of individual stages or operations. This approach could be beneficial when some 
of the stages are substantially more expensive than the other stages. This type 
of instantiation is illustrated at the bottom of the figure. In this example, the
output of the computation performed by the instances of the stage A (stage
instances A.1 and A.2) in different input data partitions is used as input to
stage instances B.1 and C.1. In our implementation, the Manager can support
both types of instantiations. 

The granularity of tasks assigned to Worker nodes is equal to stage instances,
i.e., (input data chunk, processing stage) tuples. The scheduling of tasks to
Workers is carried out using a demand-driven approach.  Stage instances are
assigned to Workers for execution in the same order the instances are created,
and the Workers continuously request work to execute as they finalize the
execution of the previous instance (see Figure~\ref{fig:execModel}). 
In practice, a single worker may execute multiple application stages concurrently, 
and the sets of Workers shown in Figure~\ref{fig:execModel} are not necessarily 
disjoint. Any necessary interprocess communication is done using 
MPI~\cite{mpi}.

Since a Worker may use a number of CPU cores and GPUs, it may ask for 
multiple stage instances from the Manager in order to 
keep all computing devices busy. The maximum number of
stage instances assigned to a Worker at a time is a configurable 
value (\emph{Window size}). The Worker may request multiple stage 
instances in one request or in multiple requests; in the latter case, 
the assignment of a stage instance and the retrieval of necessary input data 
chunks can be overlapped with the processing of an already assigned stage 
instance.   
\begin{figure}[ht]
\begin{center}
\includegraphics[width=0.44\textwidth]{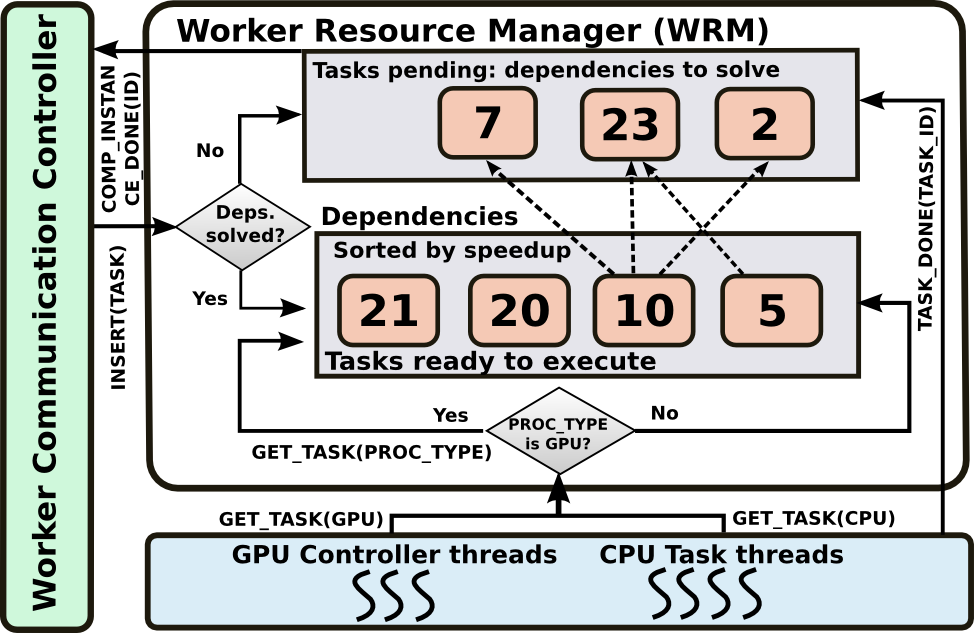}
\vspace*{-2ex}
\caption{A Worker is a multi-thread process. It uses all the devices in a hybrid node 
via the local Worker Resource Manager, which coordinates the scheduling and mapping of 
operation instances assigned to the Worker to CPU cores and GPUs.}
\label{fig:workerEnv}
\end{center}
\vspace*{-2ex}
\end{figure}

Workers  are implemented as multithread processes (see Figure~\ref{fig:workerEnv}). 
Each Worker is capable of utilizing all computing devices available within a single 
node. The Worker Communication Controller
(WCC) module runs on one of the CPU cores and is responsible for 
performing any necessary communication with the Manager. All computing 
devices used by a Worker are controlled by a local Worker Resource 
Manager (WRM).  When a Worker receives a
stage instance from the Manager and if the stage instance is composed of a pipeline of 
finer-grain operations, the Worker instantiates each of the operations in the form of 
(input data, operation) tuples, and 
dispatches the tuples to the local WRM for execution -- if the stage instance 
is a single operation, it is dispatched to the WRM as if it were a single step 
pipeline. The WRM maps the (input data, operation) tuples to the 
local computing devices as the dependencies between the operations are resolved. 
In this model of a Worker, one
computing thread is assigned to manage each available CPU computing core or a
GPU. The threads notify the WRM whenever they become idle. The WRM then selects 
one of the tuples ready for execution for that particular thread. The function 
variant is used at this point to select the appropriate operation implementation 
based on the type of the computing device.  

When all the operations in the pipeline related to a
given stage instance are executed, a callback function is invoked to notify
the WCC. The WCC then notifies the Manager about the end of that stage
instance and requests more stage instances.

\section{Runtime Optimizations} \label{sec:opts}
The baseline approach used by the WRM to decide which tuple should be 
executed next is based on a First-Come-First-Served (FCFS) approach. In this approach, the 
WRM maintains a FIFO queue of tuples. It selects the next tuple to be 
executed from the head of the queue and assigns it to the next available computing 
device. The Worker process adds to this queue new tuples as it receives more work 
from the Manager. In this section, we present several runtime optimizations that 
improve on this baseline strategy. 
\subsection{Architecture Aware Threads placement} \label{sec:tp}
\begin{figure}[htb!]
\begin{center}
\includegraphics[width=0.44\textwidth]{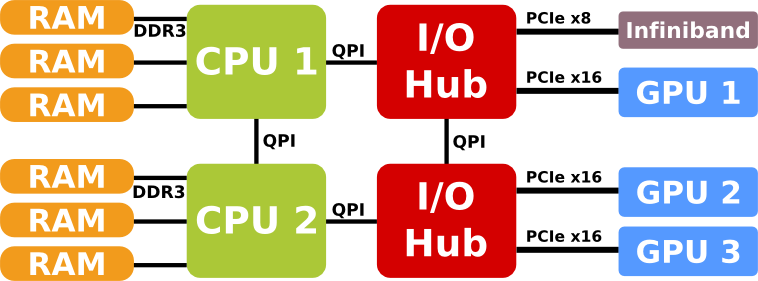}
\vspace*{-2ex}
\caption{Architecture of a Keeneland node.}
\vspace*{-2ex}
\label{fig:node}
\end{center}
\end{figure}

Machines with multiple multi-core CPUs and multiple GPUs may have heterogeneous
configurations of data paths between CPUs and GPUs to reduce
bottlenecks in data transfers between these devices. An example is the
Keeneland system~\cite{10.1109/MCSE.2011.83} used in our experimental
evaluation. Each node in Keeneland is built using three GPUs and two multi-core
CPUs, which are connected to each other through a NUMA (Non Uniform Memory
Architecture) configuration. In this configuration, there are multiple I/O
hubs, and the number of links traversed to access a GPU varies based on the CPU
used by the calling process (see Figure~\ref{fig:node}).

To use these multiple I/O hubs efficiently, it is important that CPU threads
responsible for managing GPUs be mapped to appropriate CPU cores. In our
implementation, the placement is done such that the minimum number of links is
traversed to access a given GPU. In other words, on a Keeneland node, 
CPU 1 manages GPU 1, while the thread controllers of GPU 2 and GPU 3 are 
mapped to CPU 2. This thread assignment is referred to as {\em Closest} 
in performance evaluation in the experimental results section.
\subsection{Performance Aware Task Scheduling (PATS)} \label{sec:ta}
Stage instances assigned to a Worker may create many finer-grain operation
instances. The operation instances need to be mapped to available CPU cores and
GPUs efficiently in order to fully utilize the computing capacity of a node.
Several recent efforts on task scheduling in heterogeneous environments have
targeted machines equipped with CPUs and GPUs~\cite{merge,mars,qilin09luk}.  
These works address the problem of
partitioning and mapping tasks between CPUs and GPUs for applications in which
operations (or tasks) achieve consistent and data-independent speedups when
executed on a GPU vs on a CPU. The previous efforts differ mainly in whether they 
use off-line, on-line, or automated scheduling approaches.  However, when 
there are multiple types of operations in an application, the operations may 
have different processing and data access patterns and attain different amounts 
of speedup on a GPU. Even for 
the same operation, input data characteristics may result in performance 
variability; the same operation may achieve different speedup values with
different data chunks. 

In order to use performance variability to our advantage, we have developed 
a strategy, referred here to as {\em PATS} (formerly {\em PRIORITY} 
scheduling)~\cite{Teodoro-IPDPS2012}. This strategy assigns
tasks to CPU cores or GPUs based on an estimate of the relative performance
gain of each task on a GPU compared to its performance on a CPU core and on the
computational loads of the CPUs and GPUs. We briefly describe the strategy 
here. We refer 
the reader to our earlier publication~\cite{Teodoro-IPDPS2012} for more details 
and a discussion and evaluation of performance aware task scheduling  
in the context of independent tasks. In this work, we have extended the PATS scheduler 
to take into account dependencies between operations in an analysis workflow.  

The PATS scheduler uses a queue of operation instances, i.e., {\em (data
element, operation)} tuples, sorted based on the relative speedup expected for
each tuple. As more tuples are created for execution with each Worker and
pending operation dependencies are resolved, more operations are queued for
execution. Each new operation is inserted in the queue such that the queue
remains sorted (see Figure~\ref{fig:workerEnv}). During execution, when a CPU
core or GPU becomes idle, one of the tuples from the queue is assigned to the
idle device. If the idle device is a CPU core, the tuple with the minimum
estimated speedup value is assigned to the CPU core. If the idle device is a
GPU, the tuple with the maximum estimated speedup is assigned to the GPU. The 
PATS scheduler relies on maintaining the correct relative order of speedup estimates 
rather than the accuracy of individual speedup estimates. Even if the speedup estimates 
of two tasks are not accurate with respect to their respective real
speedup values, the scheduler will correctly assign the tasks to the computing
devices on the node, as long as the order of the speedup values is correct. 
%
%
\subsection{Data Locality Conscious Task Assignment (DL)} \label{sec:dl}
GPU equipped machines are built with an extra level in the memory hierarchy, 
because discrete GPUs typically have their own memory subsystem. Input data 
for and output data from an operation may have to be transferred back-and-forth 
between CPU and GPU as operations in a pipeline are scheduled to CPUs and GPUs.
The benefits of
using a GPU for a certain computation may be strongly impacted by the cost of
data transfers between a GPU and a CPU before the GPU kernel can be started for
computation. The data transfer overheads are determined by the location where
the input data resides, and where the output data will be
stored~\cite{Gregg:2011:DWY:2015039.2015535}. 

In our execution model, input and output data are well defined as they
refer to the input and output streams of each stage and operation, which are
used by the downstream stages and operations in the analysis pipeline.
Leveraging this structure, we extend the scheduler, which in basic operation
mode uploads and downloads the input and output data used by an operation for
each assignment, with the concept of locality in order to promote data reuse
and avoid penalties due to excessive data movement. After an operation assigned 
to a GPU has finished, the scheduler explores the operation dependency graph 
and searches for operations ready for execution that can reuse the 
data already in the GPU memory. 

If the operation speedups are not known, the scheduler always chooses to reuse
data instead of selecting another operation that do not reuse data, since the
scheduler will not be able to choose a better task for GPU execution without
the speedup estimates. For the case where speedup estimates for operations are
available, the scheduler searches for tasks that reuse data in the dependency
graph, but it additionally takes into consideration other operations ready for
execution. Although those operations may not reuse data, it may be worthwhile
to pay the data transfer penalties if they benefit more from execution on a GPU
than the operations that can reuse the data. To choose which operation instance
to execute in this situation, the speedup of the dependent operation with best
speedup ($S_d$) is compared to that of the operation with the best speedup
($S_q$) in the queue that does not reuse the data.  The dependent operation is
chosen for execution, if $S_d \ge S_q \times (1-transferImpact)$. Here
$transferImpact$ is a value between 0 and 1 and represents the fraction 
of the operation execution time spent in data transfer. 
\subsection{Data Prefetching and Asynchronous Data Copy} \label{sec:dp}
Data locality conscious task assignment reduces data transfers between 
CPUs and GPUs for successive operations in a pipeline. However, there are 
moments in the execution when data still have to be exchanged between 
these devices because of scheduling decisions. In those cases, 
data copy overheads can be reduced by employing pre-fetching and 
asynchronous data copy. The typical execution of a GPU application 
involves a cyclic communication
pattern, in which data elements are copied to the GPU, the computation $kernel$ 
is launched, and output data elements are copied to the CPU memory. 
This communication pattern tends to be much slower than acyclic patterns, in which 
data can be copied to the GPU in parallel to the execution of the computation $kernel$ 
on a previously copied data~\cite{Jablin:2011:ACC:1993498.1993516}. In a similar way, 
results from previous computations may be copied to the CPU in parallel to
a $kernel$ execution. In order to employ both data prefetching and asynchronous 
data copy, we modified the runtime system to perform the computation and communication 
of pipelined operations in parallel. The execution of each operation using a 
GPU in this execution mode involves three phases: $uploading$, $processing$, and
$downloading$. Each GPU manager thread and WRM pipeline multiple operations through 
these three phases. Any input data needed for another operation waiting to execute 
and the results from a completed operation are copied from and to the CPU in 
parallel to the ongoing computation in the GPU.

\section{Experimental Evaluation} \label{sec:results}
\subsection{Experimental Setup}
We have evaluated the proposed runtime system and
optimizations using a distributed memory hybrid cluster, called
Keeneland~\cite{10.1109/MCSE.2011.83}. Keeneland is a National Science
Foundation Track2D Experimental System and has 120 nodes in the current
configuration. Each computation node is equipped with a dual socket Intel X5660
2.8 Ghz Westmere processor, 3 NVIDIA Tesla M2090 (Fermi) GPUs, and 24GB of DDR3
RAM (See Figure~\ref{fig:node}). The nodes are connected to each other through
a QDR Infiniband switch.
\begin{table}
\begin{footnotesize}
\begin{tabular}{l l l}
\hline
        Pipeline operation   	& CPU source							& GPU source 			\\ \hline \hline
\multirow{2}{*}{RBC detection}	& OpenCV and Vincent~\cite{Vincent93morphologicalgrayscale}	& \multirow{2}{*}{Implemented}	\\
				& Morph. Reconstruction (MR)          				&                             	\\ \hline
        Morph. Open             & OpenCV (by a 19x19 disk) 	  				& OpenCV               		\\ \hline
        ReconToNuclei           & Vincent~\cite{Vincent93morphologicalgrayscale} MR		& Implemented               	\\ \hline
        AreaThreshold           & Implemented   						& Implemented        		\\ \hline
        FillHolles              & Vincent~\cite{Vincent93morphologicalgrayscale} MR    		& Implemented                   \\ \hline
\multirow{2}{*}{Pre-Watershed}  & Vincent~\cite{Vincent93morphologicalgrayscale} MR and OpenCV 	& \multirow{2}{*}{Implemented}	\\
        			& for distance transformation	        			& 				\\ \hline
        Watershed            	& OpenCV               						& Korbes~\cite{Korbes:2011:AWP:2023043.2023072} \\ \hline
	BWLabel			& Implemented							& Implemented   \\ \hline
\multirow{2}{*}{Features comp.} & Implemented. 							& Implemented. 		 	\\
				& OpenCV(Canny) 						& OpenCV(Canny) 		\\
\hline
\end{tabular}
\end{footnotesize}
\vspace*{-2ex}
\caption{Sources of CPU and GPU implementations of operations in the segmentation and feature computation stages.}
\label{tab:pipeline-ops}
\vspace*{-8ex}
\end{table}

We used the example application described in Section~\ref{sec:app} for performance evaluation. 
The segmentation 
and feature computation stages were implemented as a coarse-grain workflow with two 
levels. The stages formed the first level, while the pipeline of operations in each 
stage formed the second level. As is described in Section~\ref{sec:app}, each operation 
has a CPU version and a GPU version. Several of the compute intensive operations along 
with the sources of the CPU/GPU implementations are listed in Table~\ref{tab:pipeline-ops}.
We used existing implementations from OpenCV or 
from other research groups, or implemented our own if no efficient implementations 
were available. The Morphological Open operation, for example, is
available as part of OpenCV~\cite{opencv_library} that uses NVidia Performance
Primitives (NPP)~\cite{npp}. The Watershed operation, on the other hand, has only 
a CPU implementation in the OpenCV library. We used the GPU implementation by Korbes et. al.~\cite{Korbes:2011:AWP:2023043.2023072} for this operation. We should note that the 
internal algorithms used by OpenCV and Korbes' implementations are not the same; hence, 
the results from the CPU and GPU implementations are slightly different.  
Several of the operations in the segmentation stage are irregular
computations. The Morphological Reconstruction (MR) algorithm, which is used in
these operations, has a fast CPU implementation using wave-propagation based
computation approach~\cite{Vincent93morphologicalgrayscale}. We have implemented 
a hierarchical queue-based wave-propagation framework to accelerate the MR algorithm, and 
the operations that use MR, on a GPU. The details of the framework is available as a
technical report~\cite{Teodoro-2012-Morph}. The queue-based implementation resulted
in significant performance improvements over previously published versions of
the MR algorithm~\cite{DBLP:conf/memics/Karas10}. Operations in the feature computation stage implement a number of pixel/neighborhood
based transformations that are applied to the input image (Color deconvolution,
Canny, and Gradient). Object features are extracted from the results
of the computations, based on object boundaries determined in the segmentation 
stage. Object feature computations are generally more regular and compute intensive 
than the operations in the segmentation stage. This characteristics of the feature 
computation operations lead to better GPU acceleration.

Image datasets used in the evaluation were obtained from studies in the In
Silico Brain Tumor Research Center~\cite{insilico}. Each image was partitioned
into tiles of 4K$\times$4K pixels. The codes were compiled using ``gcc 4.1.2'',
``-O3'' optimization flag, OpenCV 2.3.1, and NVIDIA CUDA SDK 4.0.  The
experiments were repeated 3 times. The standard deviation in performance
results was not observed to be higher than 2\%. The input data were 
stored in the Lustre filesystem, which is shared among multiple users. 

%
\subsection{Performance of Application Operations on GPU} 
\label{sec:comp-performance}
\begin{figure}[htb!]
\begin{center}
\includegraphics[width=0.49\textwidth]{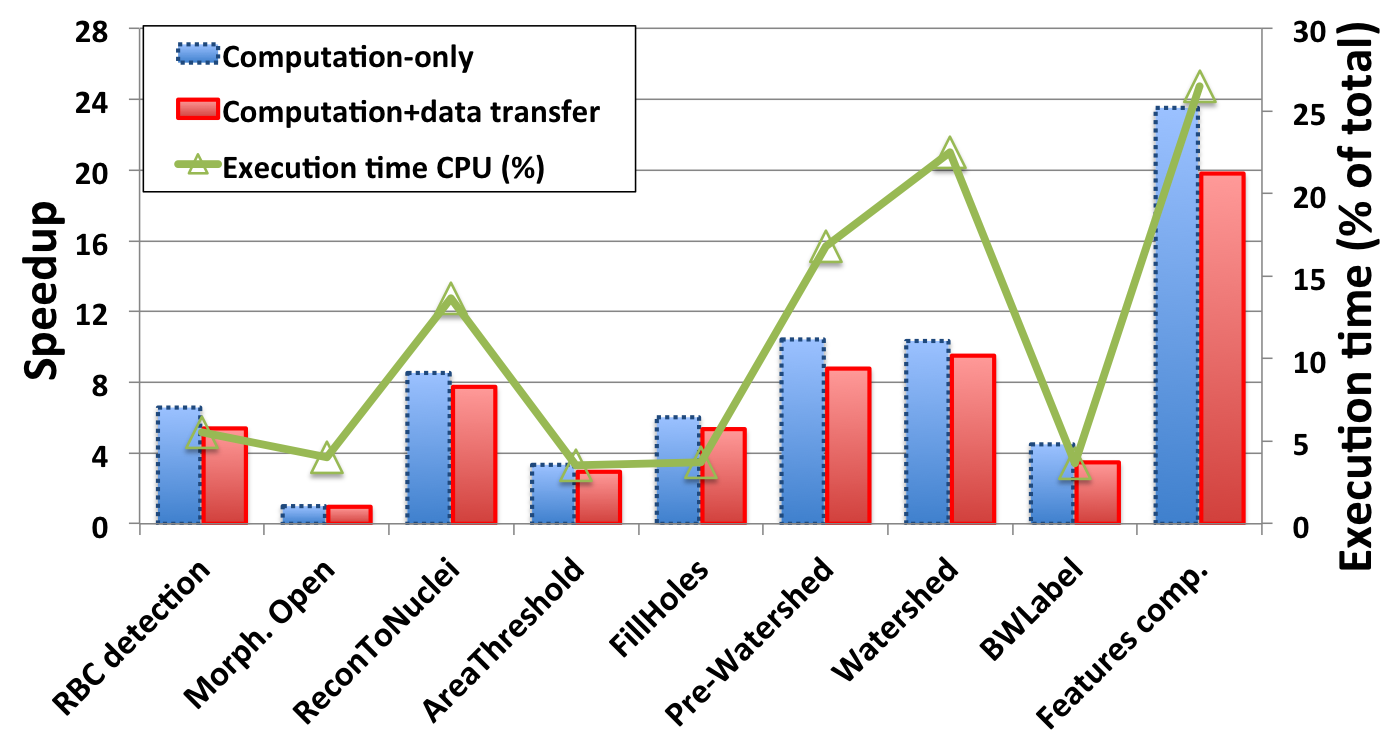}
\vspace*{-5ex}
\caption{Evaluation of the GPU-based implementations of application components (operations).}
\vspace*{-3ex}
\label{fig:componentsGPU}
\end{center}
\end{figure}

This section presents the performance gains on GPU of the
pipeline operations. Figure~\ref{fig:componentsGPU} shows the performance 
gains achieved by each of the fine-grain operations in the second level of 
the pipeline, as compared to the single core CPU counterpart. The speedup 
values in the figure represent the performance gains (1) when only the 
computation phase is considered (computation-only) and (2) when the cost 
of data transfer between CPU and GPU is included 
(computation+data transfer). The figure also shows the percentage 
of the overall computation time spent in an operation on one CPU core.  

The results show that there are significant variations in performance gains 
among operations, as expected. The most time consuming stages are 
the ones with the best speedup values -- this is in part because of the fact 
that we have focused on optimizing the GPU implementations of those operations 
to reduce overall execution time. The feature computation stage stands out as 
having better GPU acceleration than the segmentation stage. This is a 
consequence of the former stage's more regular and compute intensive nature. 

Our evaluation indicates that the task scheduling approach should take into 
consideration these performance variations to maximize performance on hybrid 
CPU-GPU platforms. We evaluate the performance impact on pipelined execution 
of using PATS for scheduling operations in Section~\ref{sec:cpu-gpu-base}.
\subsection{Architecture Aware Placement of Control Threads}
We employed two strategies for placement of CPU threads responsible for 
managing the GPUs on a cluster node. The strategies used are: (i)~OS: refers
to the placement chosen automatically by the Operation System; and,
(ii)~Closest: binds a CPU thread managing a GPU closest to that GPU regarding
to the number of connections that need to be traversed to access the GPU
(Section~\ref{sec:tp}). Three randomly selected images were used as input in the 
experiments. Each image contains 56K$\times$56K pixels and is partitioned into 
196 4K$\times$4K-pixel image tiles. Tiles with background only pixels were  
discarded beforehand, resulting in about 100 tiles per image. The image tiles 
are stored in files. The speedup results presented in this section include the 
time taken to read the input data from the Lustre filesystem.
\begin{figure}[htb!]
\begin{center}
        \includegraphics[width=0.47\textwidth]{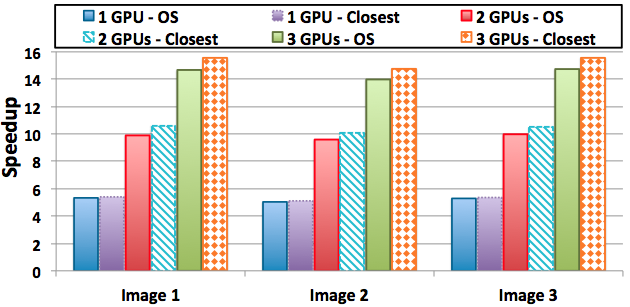}
\vspace*{-2ex}
\caption{Speedups on end-to-end execution using multiple GPUs and 
different control thread placement strategies. The results include disk I/O  
overheads to read image tiles.}
\vspace*{-2ex}
\label{fig:multi-gpu-scalability}
\end{center}
\end{figure}

The performance impact of the two strategies when 
the number of GPUs is varied as shown in Figure~\ref{fig:multi-gpu-scalability}.
The end-to-end acceleration of the pipeline using a single GPU is about
5.3$\times$ for both placement strategies, as compared to the single core CPU
version. Although a number of operations in the pipeline achieve
better speedups, the ones with lower speedups will limit the overall
gains. For instance, Morphological Open accounts for about 4\% of the CPU
execution time, but it represents about 23\% of the computation time for the
GPU accelerated version. In addition, the results also
include time spent in reading the input data which is another performance
limiting factor. If only the computation phase were considered, the GPU speedups would
be about 1.22$\times$ higher than what is reported (i.e., about 6.5$\times$ better 
performance compared to one CPU core).
 
Figure~\ref{fig:multi-gpu-scalability} also shows the performance of the
pipeline in multi-GPU configurations. The appropriate assignment of threads
managing the GPUs --- {\em Closest} --- achieved similar or superior performance than
the {\em OS} for all experiments. The {\em Closest} placement gains are higher as the
number of GPUs increases: about 3\%, 6\%, and 8\% better performance 
for the 1-, 2-, and 3-GPU configurations, respectively, in comparison to {\em OS}. This
performance is consistent across the experiments done using each of the three images. 
The Closest placement is used in the rest of the experiments
involving GPUs.
\subsection{Pipeline Execution using CPUs and GPUs in Coordination} 
\label{sec:cpu-gpu-base}
This section presents the experimental results when multiple CPU cores and GPUs are 
used together to execute the analysis workflow.   
%
%
%
In these experiments, two versions of 
the application workflow are used:
(i)~\emph{pipelined} refers to the version described in
Section~\ref{sec:app}, where the operations performed by the application are
organized as a hierarchical pipeline; (ii)~\emph{non-pipelined} that bundles the entire
computation of an input tile as a single monolithic task, which is executed
either by CPU or GPU. The comparison between these versions is important to
understand the performance impact of pipelining application operations. 

Two scheduling strategies were employed for mapping tasks to CPUs or GPUs: 
(i)~FCFS
which does not take performance variation into consideration; and, (ii)~PATS
that uses the expected speedups achieved by an operation in the scheduling
decision. When
PATS is used, the speedup estimates for each of the operations are those
presented in Figure~\ref{fig:componentsGPU}.
\begin{figure}[h!]
\begin{center}
        \includegraphics[width=0.49\textwidth]{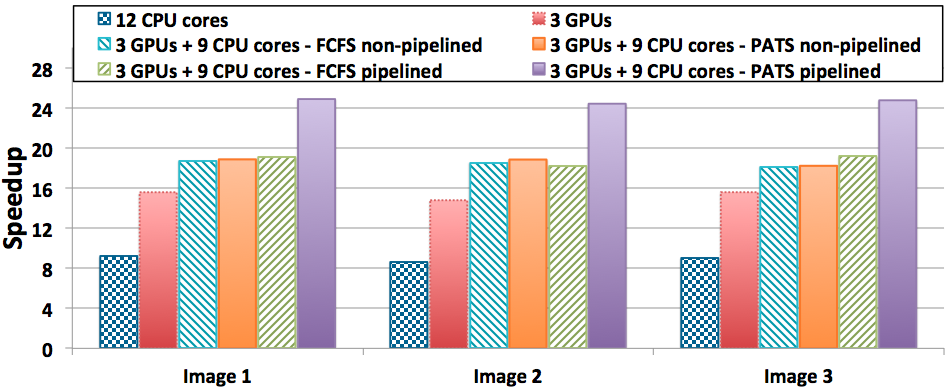}
\vspace*{-2ex}
\caption{Application scalability when multiple CPUs and GPUs are used via the PATS and FCFS 
scheduling strategies.}
\vspace*{-2ex}
\label{fig:sched-cpu-gpu}
\end{center}
\end{figure}

The results for the various configurations are presented in
Figure~\ref{fig:sched-cpu-gpu}, using  the three images from the experiments in
the previous section.  In all cases, the CPU speedup using 12 cores is about 9.
The sub-linear speedups are a result of the application's high memory
bandwidth requirements. The 3-GPU execution
achieved near linear scalability for all images. The coordinated use
of CPUs and GPUs improved performance over the 3-GPU executions. We should 
note that only upto 9 CPU cores are used in the multi-device experiments, because 
3 cores are dedicated to GPU control threads. In the non-pipelined version of 
the application, potential performance gains by using
CPUs and GPUs together are limited by load imbalance. If a tile is assigned
to a CPU core near the end of the execution, the GPUs will sit idle waiting
until the CPU core finishes, which reduces the benefits of cooperated use of
computing devices. The performance of PATS for the non-pipelined version is
similar to FCFS. In this case, the PATS scheduling is not able
to make better decisions than FCFS, because the non-pipelined version bundles 
all the internal operations of an application stage as a single task, hence the 
performance variations of the operations are not exposed to the runtime system.

The CPU-GPU execution of the pipelined version of the application with FCFS (3
GPUs + 9 CPU cores - FCFS pipelined) also improved the 3-GPU execution,
reaching similar performance to that of the non-pipelined execution. This
version of the application requires that the data are copied to and from  
a GPU before and after an operation in the pipeline is assigned
to the GPU. This introduces a performance penalty due to the data transfer 
overheads, which are about 13\% of the computation time as show in 
Figure~\ref{fig:componentsGPU}, and limits the performance improvements of 
the pipelined version. The advantage of using the pipelined version in this
situation is that load imbalance among CPUs and GPUs is reduced. The assignment 
of computation to CPUs or GPUs occurs at a
finer-grain; that is, application operations in the second level of the pipeline 
make up the tasks scheduled to CPUs and GPUs, instead of the entire computation 
of a tile as in the non-pipelined version.
\begin{figure}[h!]
\begin{center}
        \includegraphics[width=0.49\textwidth]{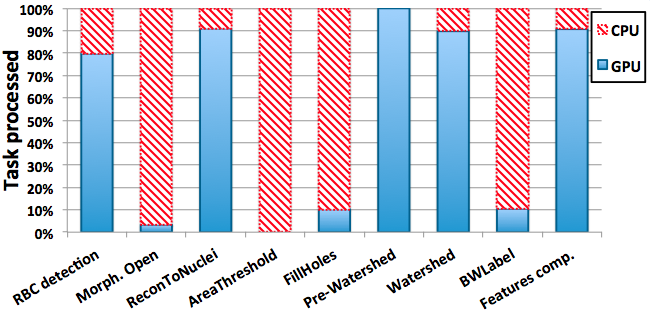}
\vspace*{-5ex}
\caption{Execution profile (\% of tasks processed by CPU or GPU) using PATS per pipeline stage.}
\vspace*{-2ex}
\label{fig:sched-profile}
\end{center}
\end{figure}

Figure~\ref{fig:sched-cpu-gpu} also presents the performance of the PATS
scheduling for the pipelined version of the application.
As is seen from the figure, processing of tiles using PATS is about 
1.33$\times$ faster than using FCFS with the non-pipelined or pipelined 
version of the application. The performance gains result from the ability of PATS 
to better assign the application internal operations to the
most suited computing devices. For instance, Figure~\ref{fig:sched-profile} presents
the percent of tasks that PATS assigned to the CPUs or GPUs for each pipeline stage. 
As is shown, the execution of components with lower speedups are mostly performed 
using the CPUs, while the GPUs are kept occupied with
operations that achieve higher speedups. For reference, using FCFS with the
pipelined version, the operations are more or less evenly distributed across 
CPUs and GPUs regardless of performance variations between the operations. 
\subsection{Data Locality Conscious Scheduling and Data Prefetching}
\label{sec:cpu-gpu-dl-p}
\begin{figure}[h!]
\begin{center}
        \includegraphics[width=0.5\textwidth]{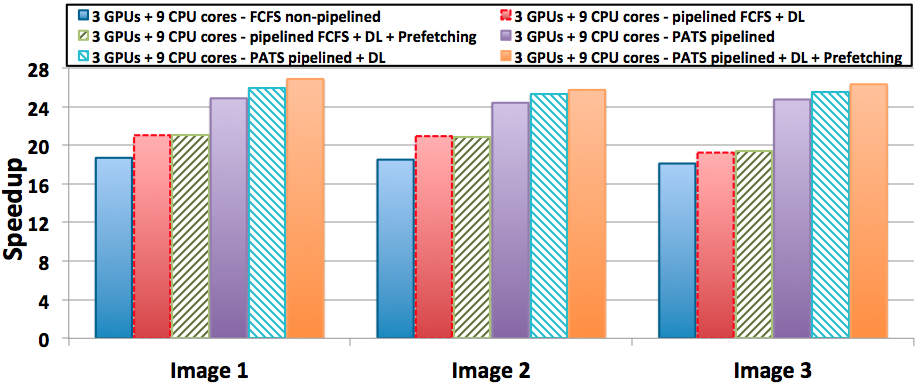}
\vspace*{-2ex}
\caption{Performance impact of data locality conscious mapping and asynchronous data copy 
optimizations.}
\vspace*{-2ex}
\label{fig:sched-cpu-gpu-opts}
\end{center}
\end{figure}

In this section, we evaluate the performance impact of the data locality conscious task
assignment (DL) and data prefetching and asynchronous data download
(Prefetching) optimizations. Figure~\ref{fig:sched-cpu-gpu-opts} presents the
performance improvements with these optimizations for both PATS and FCFS policies.
As is shown, the pipelined version with FCFS and DL is able to improve the
performance of the non-pipelined version by about 1.1$\times$ for all input 
images. When Prefetching is used in addition to FCFS and DL (``3GPUs + 9 CPU core -
pipelined FCFS + DL + Prefetching''), there are no significative performance
improvements. The main reason is that DL already avoids any unnecessary CPU-GPU
data transfers; therefore, Prefetching will only be effective in reducing
the cost of uploading the input tile to the GPU and downloading the final results 
from the GPU. These
costs are small and limit the performance gains resulting from Prefetching.

Figure~\ref{fig:sched-cpu-gpu-opts} also shows the performance results for
PATS when DL and Prefetching are employed. The use of DL improves the
performance of PATS as well, but the gains achieved (1.04$\times$) 
with DL are smaller than those in FCFS. In this case, the estimated speedups for the
operations are available, thus PATS will check whether it is worthwhile to
download the operation results to map another operation to the GPU. 
The number of upload/downloads avoided by using DL is
also smaller than when FCFS is used, which explains the performance gain 
difference. Prefetching with DL results in an additional 1.03$\times$ 
performance improvement. This
optimization was more effective in this case because the volume of data
transferred between the CPU and the GPU is much higher than when FCFS with DL is 
employed.
\subsection{Impact of Worker Request Window Size}
This section analyzes the effect of the demand-driven window size between
Manager and Workers (i.e., the number of pipeline stage instances  
concurrently assigned to a Worker) on the CPU-GPU scheduling 
strategies utilized by the Worker. During
this evaluation, we used 3 GPUs and 9 CPU cores (with 3 CPU cores allocated to 
the GPU manager threads) with FCFS and PATS. The window-size is varied from 12 
until no significant performance changes are observed.

\begin{table}[!h]
\begin{small}
\begin{tabular}{c c c c c c c c c}
\hline
	& \multicolumn{8}{c}{Demand-Driven Window Size}				\\ \cline{2-9}
       	& 12   	& 13	& 14 	& 15	& 16	& 17	& 18	& 19 	\\ \hline
FCFS	& 75.1  & 73.4	& 74.9 	& 73.7	& 75.3	& 74.9	& 73.2	& 73.5	\\ \hline
PATS 	& 75.1  & 61	& 56.9 	& 53.1	& 54.1	& 51.5	& 51.2	& 50.7	\\ \hline
\end{tabular}
\end{small}
\vspace*{-2ex}
\caption{Execution time (secs.) for different request window size and scheduling policies using 3 GPUs and 9 CPU cores.}
\label{tab:demand-driven-time}
\vspace*{-6ex}
\end{table}


Table~\ref{tab:demand-driven-time} presents the execution times. 
FCFS scheduling is impacted little by variation in the window size. The PATS
scheduler performance, on the other hand, is limited for small window sizes.
In the scenario where the window size
is 12, FCFS and PATS tend to make the same scheduling decisions, because
only a single operation will be available when a processor requests work. This 
makes the decision trivial and equal for both strategies. When the window size is
increased, however, the scheduling decision space becomes larger, providing
PATS with opportunities to make better task assignments. As is shown in the table, 
with a window size of 15, PATS already achieves near its best performance. This is
another good property of PATS, since very large window sizes
can create load imbalance among Workers.

\begin{figure}[htb!]
\begin{center}
        \includegraphics[width=0.46\textwidth]{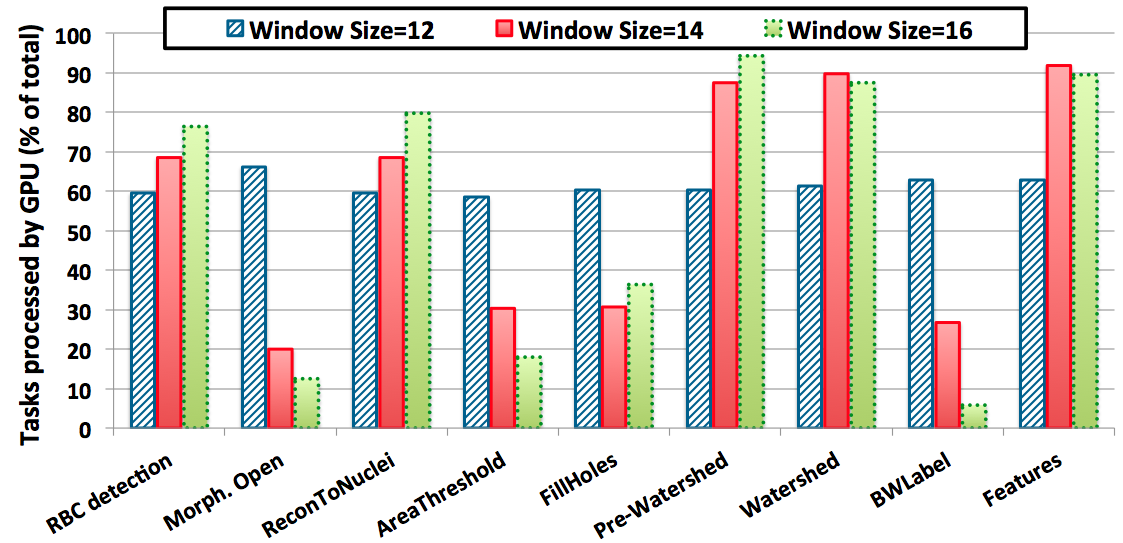}
\vspace*{-2ex}
\caption{Execution scheduling profile for different window sizes and the PATS strategy.}
\vspace*{-3ex}
\label{fig:demand-driven-profile}
\end{center}
\end{figure}

The profile of the execution (\% of tasks processed by GPU) as the window size 
is varied is displayed in Figure~\ref{fig:demand-driven-profile}. As the window 
size increases, PATS quickly changes the assignment of tasks, and
operations with higher speedups are more likely to be executed by GPUs.  FCFS
profile is not presented in the same figure, but its profile is similar to PATS 
with a window size of 12 for all configurations.
\subsection{Effects of Inaccurate Speedup Estimation}

In this section, we empirically evaluate the sensitivity of PATS scheduler to
errors in the GPU-vs-CPU speedup estimation of operations. For the sake of this 
analysis, we intentionally inserted errors in the estimated speedup values of 
the application operations in a controlled manner. In
order to effectively confound the method, operations with lower speedups that
are mostly scheduled to the CPUs (Morph. Open, AreaThreashold, FillHoles, and
BWLabel; see Figure~\ref{fig:sched-profile}) had their estimated speedup 
values increased, while the others have the values decreased.  The changes
are calculated as a percentage of an operation's original estimated speedup,
and the variation range was from 0\% to 100\%.

\begin{figure}[htb!]
\begin{center}
        \includegraphics[width=0.47\textwidth]{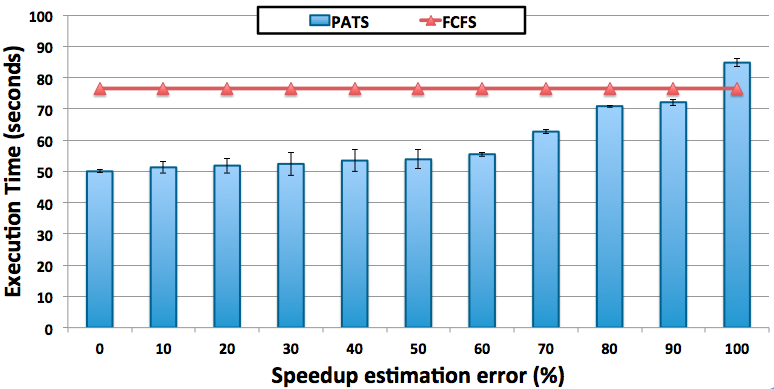}
\vspace*{-2ex}
\caption{Performance of PATS when errors in speedup estimation for the pipeline operations 
are introduced.}
\vspace*{-3ex}
\label{fig:estimation-error}
\end{center}
\end{figure}

The execution times for different error rates are presented in
Figure~\ref{fig:estimation-error}. The results show that PATS
is capable of performing well even with high errors and error rates in 
speedup estimations. For instance, when 60\% estimation error is used, the performance of the
pipeline is only 10\% worse than the initial case (0\% speedup estimation
error). At 70\% and 80\% errors, PATS performance is more impacted, as
a result of a miss-ordering of the pipeline operations before mostly
processed by CPU (AreaThreashold, FillHoles, and BWLabel) with ReconToNuclei
and Watershed. Consequently, those stages with lower speedups will be scheduled 
for execution on a GPU. Nevertheless, PATS still performs better than FCFS, 
because the operations in the feature computation stage are not miss-ordered. 
To emulate 100\% estimation error, we set to 0 the speedups of all substages that 
in practice have higher speedups, and double the estimated speedups of the other 
stages that in reality have lower speedup values. This forces PATS to preferably 
assign operations with low speedups to GPU and the ones with high speedup to CPU. 
Even with this level of 
error, the execution times are only about 
10\% worse than those using FCFS.
\subsection{Multi-node Scalability}
\begin{figure}[htb!]
\begin{center}
        \includegraphics[width=0.47\textwidth]{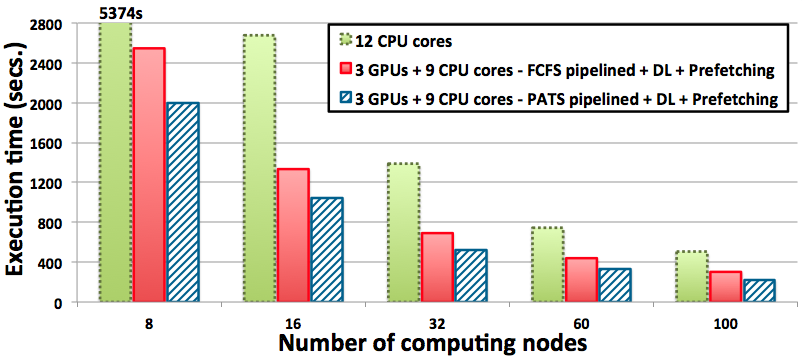}
\vspace*{-2ex}
\caption{Multi-node scalability: strong scaling evaluation.}
\vspace*{-4ex}
\label{fig:strong-scale}
\end{center}
\end{figure}

This section presents the performance evaluation of the runtime system when multiple 
computation nodes are used. The evaluation was carried out using 340 glioblastoma 
brain tumor WSIs, which were partitioned into a total of 36,848 4K$\times$4K tiles. 
Figure~\ref{fig:strong-scale} shows the execution times for all configurations when 
the number of computing nodes is varied from 8 to 100.  As in the other experiments, 
the input tiles were stored in the Lustre filesystem. 
As is shown in the figure, PATS with the other optimizations achieved the best 
performance with a speedup of 1.3$\times$ over FCFS. The efficiency of the runtime 
system on 100 nodes is about 77\%. The main limiting factor and bottleneck is the I/O 
overhead of reading image tiles. As the number of nodes increases, I/O operations become
more expensive, because more clients access the file system in parallel. If only the computation 
times were measured, the efficiency would increase to about 93\%. Even with the I/O overheads, 
the runtime system was able to process the entire
set of 36,848 tiles in less than four minutes when 100 nodes were used. This 
represents a huge improvement in computing capabilities; the same computation
would take days or weeks on a workstation using the original MATLAB 
version. 

%

\section{Conclusions}
Hybrid CPU-GPU cluster systems offer significant computing and memory capacity to 
address the computational needs of large scale scientific analyses. We have developed 
a middleware system to tap this capacity and enable high-throughput data processing by 
leveraging common data access and processing patterns in scientific analysis 
applications. Findings from the experimental evaluation of this system on a 
state-of-the-art hybrid cluster system can be summarized as follows: Significant 
performance improvements can be achieved when an analysis application can be assembled 
as pipelines of fine-grain operations compared to bundling all internal operations 
in one or two monolithic methods. The former allows for exporting application 
processing patterns more accurately to the runtime environment and empowers 
the middleware system to make better scheduling decisions. Performance aware 
task scheduling coupled with function variants enable efficient coordinated use of 
CPU cores and GPUs in pipelined operations. Performance gains can further be 
increased on hybrid systems through such additional runtime optimizations as 
locality conscious task mapping, data prefetching, and asynchronous data copy. 
Employing a combination of these optimizations, our runtime system implementation 
has achieved a processing rate of about 150 tiles per second when 100 nodes are 
used. These levels of processing speed make it feasible to process very large 
datasets and would enable a scientist to explore different scientific questions 
rapidly and/or carry out algorithm sensitivity studies. 

The work presented in this paper has focused on the segmentation and feature
computation stages of the example analysis application. We plan to extend
support to the classification stage. This stage implements a MapReduce style
processing pattern. Moreover, although the classification stage in the current
application implementation is relatively inexpensive, since it operates on
aggregated image and patient level data, there are plans to extend it to
support clustering-based classifications using object level data -- the number
of segmented nuclei in a large data set can reach hundreds of millions, even
billions. We are currently in the process of developing fast CPU and GPU
implementations for clustering large volumes of point data. We plan to
integrate these function variants along with support for MapReduce type of
processing in order to provide full support for analysis applications that are
similar to the example analysis application. 

\noindent {\bf Acknowledgments.} This work was supported in part by
HHSN261200800001E from the National Cancer Institute,
R24HL085343 from the National Heart Lung and Blood Institute, by 
R01LM011119-01 and R01LM009239 from the National Library of Medicine,
RC4MD005964 from National Institutes of Health,
and PHS UL1RR025008 from the Clinical and Translational Science Awards
program. This research used resources of the Keeneland Computing Facility at the 
Georgia Institute of Technology, which is 
supported by the National Science Foundation under Contract OCI-0910735.

\balance

%
\bibliographystyle{IEEEtran}
\bibliography{george}  
%

\end{document}